\documentclass[12pt]{article}
\usepackage{amsfonts}
\usepackage{amsmath}
\usepackage{amssymb}
\usepackage{mathptmx}
\usepackage{geometry}
\usepackage{graphicx}
\usepackage{hyperref}
\usepackage{cancel}
\usepackage{textcomp}
\usepackage{tikz}
\usepackage{bm}
  \usepackage{mathrsfs}
  \usepackage{filecontents}
\usepackage{times}
\usepackage{epsfig}
\usepackage{xcolor}
\usepackage{slashed}
\usepackage{booktabs}
\usepackage{latexsym}
\usepackage{verbatim}
\usepackage{extarrows}
\usepackage{multirow}
\usepackage{rotating}
\usepackage{colortbl}
\usepackage{indentfirst}
\usepackage{float} 
\definecolor{mygray}{gray}{.9}
\definecolor{intnull}{RGB}{213,229,255}
\usepackage{arydshln}
\usepackage{diagbox}
\usepackage{makecell}
\usepackage{array}

\usepackage{caption}
\usepackage{tikz}
\usetikzlibrary{arrows,shapes,chains}
\usepackage{graphicx, subfig}
\begin{document}
\renewcommand{\thefootnote}{\fnsymbol{footnote}}
\baselineskip=16pt
\pagenumbering{arabic}
\vspace{1.0cm}
\begin{center}
{\Large\sf  Quasinormal modes of  the generalized ABG STVG black hole in the scalar-tensor-vector gravity}
\\[10pt]
\vspace{.5 cm}

{Xin-Chang Cai\footnote{E-mail address: caixc@mail.nankai.edu.cn} and Yan-Gang Miao\footnote{Corresponding author. E-mail address: miaoyg@nankai.edu.cn}}
\vspace{3mm}

{School of Physics, Nankai University, Tianjin 300071, China}

\vspace{4.0ex}

{\bf Abstract}
\end{center}

We obtain the solution of a generalized ABG STVG black hole with the nonlinear tensor field $B_{\mu \nu }$ in the scalar-tensor-vector gravity. This black hole is endowed with four parameters, the black hole mass $M$, the parameter $\alpha$ associated with the STVG theory, and the two parameters $\lambda$ and $\beta$ that are related to the dipole and quadrupole moments of the nonlinear tensor field, respectively. By analyzing the characteristics of the black hole, we find that  the generalized ABG STVG black hole is regular when $\lambda \geqslant 3$ and  $\beta \geqslant 4$, and  we study the effects of the  parameters  $\alpha$, $\lambda$ and $\beta$ on the black hole horizon. We calculate the quasinormal mode frequencies of the odd parity gravitational perturbation for the generalized ABG STVG black hole by using the 6th order WKB approximation method and simultaneously by the null geodesic method at the eikonal limit. The results show that the increase of the parameters $\alpha$ and $\lambda$ makes the gravitational waves decay slowly, while the increase of the parameter $\beta$ makes the gravitational waves decay fast at first and then slowly. In addition, we verify that the improved correspondence between  the real part of quasinormal frequencies at the eikonal limit and the shadow radius is valid for the generalized ABG STVG black hole.

\newpage

\section{Introduction}
The gravitational waves from binary black hole mergers detected by the LIGO and Virgo collaborations~\cite{P12,P13} have stimulated our interest in black holes. The nature of the detected gravitational wave signals is the quasinormal mode with a complex frequency, and the main composition is its fundamental mode with the lowest frequency. This complex frequency is called a quasinormal mode frequency whose real part and imaginary part represent the oscillating and damping of the gravitational wave, respectively. The quasinormal mode is determined only by the parameters of a black hole itself. Due to its importance in the study of black holes, the quasinormal modes of various classical or semiclassical black holes have been analyzed  extensively and deeply in Einstein's gravity and beyond in the past few decades, see, for instance, some reviews~\cite{P14,P15,P16,P17}.

Although Einstein's theory of gravity has achieved great success and withstood many experimental tests in the past 100 years, it cannot explain~\cite{P17,P18,P19,P20,P21} the discrepancy between the dynamics of galaxies and the amount of luminous matter contained in the galaxies. The exotic dark matter was introduced~\cite{P22} to explain this discrepancy. However, the dark matter has not yet been detected so far. An alternative way to solve the problem of the galaxy dynamics was to modify~\cite{P22} the laws of gravity on the scales that have not been extensively tested by Newtonian gravity or general relativity. In various versions of modified gravity,
the scalar-tensor-vector gravity (STVG) theory~\cite{P1} is a good candidate, which contains a scalar action, a vector action and a matter action. The STVG theory is able to describe~\cite{P23,P24,P25,P26} the dynamics of galaxies without assuming the existence of dark matter in the universe. In addition, this theory has been applied to deal with the problems in the solar system, such as the growth of structure~\cite{P27}, and in the early universe, for instance, the cosmic microwave background (CMB) acoustical power spectrum data~\cite{P28}, and so on. Therefore, it is of great significance to compute the quasinormal modes of black holes in the STVG theory for the analysis of gravitational waves. To this end, the quasinormal modes of the electromagnetic and gravitational perturbations~\cite{P29,P42} of STVG black holes have been studied, as well as the quasinormal modes of STVG black holes at the eikonal approximation~\cite{P33}.

It is well known that the existence of the singularity at the center of a black hole will cause all the laws of physics to fail there.  In order to solve the problem of curvature singularity, Bardeen proposed~\cite{P30} the first regular black hole solution, now called the Bardeen black hole, which has been successfully interpreted~\cite{P31} by Ay\'{o}n-Beato and Garc\'{\i}a (ABG) as a magnetic solution to Einstein's equations coupled to nonlinear electrodynamics. In addition, Ay\'{o}n-Beato and Garc\'{\i}a have obtained a series of regular black hole solutions by coupling various nonlinear electrodynamics to the Einstein field equations, such as the  ABG black hole~\cite{P32} and the generalized ABG black hole~\cite{P3}. Following Ay\'{o}n-Beato and Garc\'{\i}a's method, Moffat has given~\cite{P2} the ABG black hole solution without singularity in the STVG theory. Considering that the ABG black hole is only a special case of the generalized ABG black hole, we think that it is meaningful to find out the generalized ABG black hole solution in the STVG theory, called the generalized ABG STVG black hole.
Furthermore, considering that its quasinormal modes are associated with the nonvanishing perturbed energy-momentum tensor rather than the vacuum case, we think that it is nontrivial to study the corresponding quasinormal modes of the odd parity gravitational perturbation. Our work may be regarded as a crucial supplement to both the regular black holes and the STVG theory, which completes the construction of black hole solutions composed of the four parts, the ABG and generalized ABG black holes in Einstein's theory, and the ABG and generalized ABG black holes in the STVG theory.

The paper is organized as follows. In Sect. 2, we briefly introduce the field equations of the STVG theory and the related Schwarzschild STVG black hole. In Sect. 3, we find out the exact solution of the generalized ABG STVG black hole. Then we analyze the characteristics of this black hole model in Sect. 4. In Sect. 5, we calculate the quasinormal mode frequencies of  the odd parity gravitational perturbation for the generalized ABG STVG black hole and the quasinormal mode frequencies at the eikonal limit as well. Finally, we make a simple summary in Sect. 6. We use the units $c=G_{N}=1$ and the sign convention $(-,+,+,+)$ throughout the paper.

\section{Field equation of STVG theory and Schwarzschild STVG black hole}

The  action of the STVG  theory is as follows~\cite{P1},
\begin{equation}
\label{1}
S=S_{GR}+S_{\phi }+S_{S}+S_{M},
\end{equation}
with
\begin{equation}
\label{2}
S_{GR}=\frac{1}{16\pi }\int d^{4}x\sqrt{-g}\frac{1}{G}R,
\end{equation}
\begin{equation}
\label{3}
S_{\phi }=-\frac{1}{4\pi }\int d^{4}x\sqrt{-g}\left(K-\frac{1}{2}\tilde{\mu}^{2}\phi ^{\mu }\phi _{\mu}\right),
\end{equation}
\begin{equation}
\label{4}
S_{S}=\int d^{4}x\sqrt{-g}\left[\frac{1}{G^{3}}\left(\frac{1}{2}g^{\mu \nu }\nabla _{\mu }G \nabla_{\nu  }G-V(G)\right)+\int d^{4}x\frac{1}{\tilde{\mu}^{2}G}\left(\frac{1}{2}g^{\mu \nu }\nabla _{\mu }\tilde{\mu} \nabla_{\nu  }\tilde{\mu}-V(\tilde{\mu})\right)\right].
\end{equation}
Here, $S_{GR}$ represents the Einstein-Hilbert action, $S_{\phi }$ the action of a massive vector field, $S_{S}$ the action of a scalar field, and $S_{M}$ the action of possible matter sources. $\phi ^{\mu }$ denotes a Proca-type massive vector field with mass $\tilde{\mu}$, and $K$ the kinetic term for the vector field $\phi ^{\mu}$ usually chosen as $K=\frac{1}{4}B^{\mu \nu}B_{\mu \nu}$ with the linear tensor field $B_{\mu \nu }=\partial _{\mu }\phi _{\nu }-\partial _{\nu}\phi _{\mu  }$. $G(x)$ and $\tilde{\mu}(x)$ are two scalar fields that vary with respect to space and time, and $V(G)$ and $V(\tilde{\mu})$ their corresponding potentials, respectively.

On the one hand, the effect of the mass $\tilde{\mu}$ of the vector field $\phi ^{\mu }$ manifests at kiloparsec scales from the gravitational source, so it can be neglected when we solve the field equations for a black hole solution. On the other hand, we regard $G$ as a constant that depends on the parameter $\alpha $, i.e., $G=G_{N}(1+\alpha )$, where $G_{N}$ represents Newton's gravitational constant and $\alpha $ is a dimensionless parameter. For $\alpha =0$, the STVG theory returns to Einstein's general relativity (GR), so we can regard $\alpha $ as a deviation parameter of the STVG theory from GR. Furthermore, we can simplify Eq.~(\ref{1}) for the vacuum solution, 
\begin{equation}
\label{5}
S=\frac{1}{16\pi }\int d^{4}x\sqrt{-g}\left(\frac{R}{ G }-B^{\mu \nu}B_{\mu \nu}\right).
\end{equation}

When variating the action Eq.~(\ref{5}) with respect to $g_{\mu \nu}$, one obtains the field equation, 
\begin{equation}
\label{6}
G_{\mu \nu }=-8\pi GT_{\mu \nu }^{\phi },
\end{equation}
where $G_{\mu \nu}$ is the Einstein tensor, and the energy-momentum tensor $T_{\mu \nu}^{\phi}$ with respect to the vector field $\phi ^{\mu}$ takes the form,
\begin{equation}
\label{7}
T_{\mu \nu}^{\phi}=-\frac{1}{4\pi}\left({B_{\mu}}^{\sigma}B_{\nu \sigma}-\frac{1}{4}g_{\mu \nu}B^{\rho \sigma}B_{\rho \sigma}\right).
\end{equation}
When variating the action Eq.~(\ref{5}) with respect to the vector field $\phi^{\mu}$, one obtains the field equation,
\begin{equation}
\label{8}
\nabla _{\nu  }B^{\mu \nu }=0,
\end{equation}
\begin{equation}
\label{9}
\nabla _{\sigma }B_{\mu \nu }+\nabla _{\mu  }B_{ \nu\sigma  }+\nabla _{\nu }B_{\sigma \mu  }=0.
\end{equation}

In order to get the solution of a static spherically symmetric black hole in the STVG, the line element takes the following form, 
\begin{equation}
\label{10}
ds^{2}=-f(r)dt^{2}+\frac{dr^{2}}{f(r)}+r^{2}\left(d\theta^{2}+\sin^{2}\theta d\varphi^{2}\right).
\end{equation}
Using the assumption that the gravitational source charge $q$ of the vector field $\phi ^{\mu }$ is proportional to the source mass $M$, i.e., $q=\sqrt{\alpha G_{N}}M$, Moffat proposed~\cite{P2} the metric function of the Schwarzschild STVG black hole by solving Eqs.~(\ref{6}) and (\ref{8}), 
\begin{equation}
\label{11}
f(r)=1-\frac{2GM}{r}+\frac{\alpha G_{N}GM^{2}}{r^{2}}.
\end{equation}
Similar to the Reissner-Nordstr\"om  black hole, the Schwarzschild STVG has two horizons,
\begin{equation}
\label{12}
r_{\pm }=G_{N}M\left(1+\alpha \pm\sqrt{1+\alpha}\right),
\end{equation}
where $r_{-}$ is the inner horizon called the Cauchy horizon and $r_{+}$ is the outer horizon called the event horizon. In addition, when $\alpha =0$, the two horizons are merged into the Schwarzschild event horizon.

\section{Construction of the generalized ABG STVG black hole}

Since there is no reason that the usual linear kinetic term for the vector field $\phi ^{\mu }$ must be chosen, we can consider the nonlinear kinetic term, i.e., $K=L(P)=2PH_{P}-H(P)$, where $P\equiv \frac{1}{4}P_{\mu \nu }P^{\mu \nu }$,  $H_{P}\equiv \frac{\mathrm{d} H(P)}{\mathrm{d} P}$, and $P_{\mu \nu }\equiv \frac{B_{\mu \nu }}{H_{P}}$. In order to determine the nonlinear kinetic term, we choose the structural function $H(P)$ given in Ref.~\cite{P3} as follows,

\begin{equation}
\label{13}
H(P)=P\,\frac{1-(\beta -1)\sqrt{-2q^{2}PG^{2}}}{\left(1+\sqrt{-2q^{2}PG^{2}}\right)^{\frac{\beta}{2}+1}}-\frac{\lambda }{2q^{2}sG^{\frac{3}{2}}}\frac{\left(\sqrt{-2q^{2}PG^{2}}\right)^{\frac{5}{2}}}{\left(1+\sqrt{-2q^{2}PG^{2}}\right)^{\frac{\lambda  }{2}+1}},
\end{equation}
where $s=\frac{q}{2M}=\frac{\sqrt{\alpha G_{N}}}{2}$, the invariant $P$ is a negative  quantity, and  $\beta $ and $\lambda$ are two dimensionless parameters. In addition, the action, Eq.~(\ref{1}), now becomes the following form,
\begin{equation}
\label{14}
S=\int d^{4}x\sqrt{-g}\left(\frac{1}{16\pi }\frac{R}{ G }-\frac{1}{4\pi }L(P)\right).
\end{equation}

By variating the above action with respect to the metric $g_{\mu \nu }$ and the vector field $\phi ^{\mu }$, respectively, we derive the nonlinear STVG field equations,
\begin{equation}
\label{15}
G_{\mu }^{\nu }=2G\left[\mathit{H}_{P}P_{\mu \sigma}P^{\nu \sigma}-\delta _{\mu}^{\nu }\left(2P\mathit{H}_{P}-\mathit{H(P)}\right)\right],
\end{equation}
\begin{equation}
\label{16}
\nabla _{\mu  }P^{\gamma \mu }=0.
\end{equation}
By assuming that the metric function $f(r)=1-\frac{2M(r)}{r}$ in the line element, Eq.~(\ref{10}), and  the antisymmetric field $P_{\mu \nu }=2\delta^{t}_{[\mu } \delta_{\nu] }^{r}D(r)$, where $D(r)$ is just a function of $r$, and using Eq.~(\ref{16}), we can get
\begin{equation}
\label{17}
P_{\mu \nu }=2\delta^{t}_{[\mu } \delta_{\nu] }^{r}\frac{q}{r^{2}}   \quad  \Longrightarrow \quad P=-\frac{D^{2}}{2}=-\frac{q^{2}}{2r^{4}},
\end{equation}
where we have chosen the integration constant as the gravitational source charge $q$, i.e., $q=\sqrt{\alpha G_{N}}M$. In addition, the corresponding ${G_{t}}^{t}$ component of the STVG equations, see Eq.~(\ref{15}), now reads
\begin{equation}
\label{18}
\frac{\mathrm{d} M(r)}{\mathrm{d} r}=-Gr^{2}\mathit{H}(P).
\end{equation}
By substituting the formula Eq.~(\ref{13}) and $P=-\frac{q^{2}}{2r^{4}}=-\frac{\alpha G_{N}M^{2}}{2r^{4}}$ into Eq.~(\ref{18}) and then integrating, we obtain
\begin{equation}
\label{19}
M(r)=\frac{MGr^{\lambda }}{\left(r^{2}+\alpha G_{N}GM^{2}\right)^{\frac{\lambda }{2}}}-\frac{\alpha G_{N}GM^{2}r^{\beta -1}}{2\left(r^{2}+\alpha G_{N}GM^{2}\right)^{\frac{\beta}{2}}}+C,
\end{equation}
where $M$ is the black hole mass and $C$ the integration constant. Considering that $M(r)$ should also be equal to zero when $M=0$, we set $C = 0$. Finally, by substituting the formula Eq.~(\ref{19}) into the metric function $f(r)=1-\frac{2M(r)}{r}$, we compute the metric of the generalized ABG STVG black hole,
\begin{equation}
\label{20}
f(r)=1-\frac{2MGr^{\lambda -1}}{\left(r^{2}+\alpha G_{N}GM^{2}\right)^{\frac{\lambda }{2}}}+\frac{\alpha G_{N}GM^{2}r^{\beta -2}}{\left(r^{2}+\alpha  G_{N}GM^{2}\right)^{\frac{\beta}{2}}}.
\end{equation}

\section{Characteristics of the generalized ABG STVG black hole}

According to the metric function Eq.~(\ref{20}), we can calculate the scalar curvature of the generalized ABG STVG black hole as follows,
\begin{equation}
\label{21}
R=\frac{2\lambda \alpha G_{N}G^{2}M^{3}\left(\alpha (\lambda +1)G_{N}M^{2}G-r^{2}\right)r^{\lambda -3}}{\left(r^{2}+\alpha G_{N}GM^{2}\right)^{\frac{\lambda }{2}+2}}-\frac{\beta \alpha^{2} G_{N}^{2}G^{2}M^{4}\left(\alpha (\beta -1)G_{N}M^{2}G-3r^{2}\right)r^{\beta -4}}{\left(r^{2}+\alpha G_{N}GM^{2}\right)^{\frac{\beta }{2}+2}}.
\end{equation}
From the above formula, we can easily find that the scalar curvature is not divergent at $r=0$ and is regular everywhere when $\lambda \geqslant 3$ and  $\beta \geqslant 4$, which also means that  the generalized ABG STVG black hole is a regular black hole in this case. When $\lambda=3$ and  $\beta=4$, the metric function Eq.~(\ref{20}) becomes
\begin{equation}
\label{22}
f(r)=1-\frac{2MGr^{2}}{\left(r^{2}+\alpha G_{N}GM^{2}\right)^{\frac{3}{2}}}+\frac{\alpha G_{N}GM^{2}r^{2}}{\left(r^{2}+\alpha  G_{N}GM^{2}\right)^{2}},
\end{equation}
which is just the ABG STVG black hole given by Moffat~\cite{P2}. In addition, when $\lambda=\beta=0$, the metric function Eq.~(\ref{20}) goes back to Eq.~(\ref{11}), the (singular) Schwarzschild STVG black hole. By expanding the metric function Eq.~(\ref{20}) asymptotically until $O(\frac{1}{r^{5}})$, we have
\begin{equation}
\label{23}
1-\frac{2GM}{r}+\frac{\alpha G_{N}GM^{2}}{r^{2}}+\frac{\lambda \alpha G_{N}G^{2}M^{3}}{r^{3}}-\frac{1}{2}\frac{\beta \alpha ^{2}G_{N}^{2}G^{2}M^{4}}{r^{4}}+O(\frac{1}{r^{5}}).
\end{equation}
By analogy to the physical interpretation of the parameters $\lambda$ and $\beta$ of the generalized ABG  black hole in Einstein's gravity given by Ref.~\cite{P3}, we think that the parameters $\lambda$ and $\beta$ are associated with the dipole and quadrupole moments of   the nonlinear tensor field $B_{\mu \nu }$, respectively.

Taking $M=G_{N}=1$, the metric function~(\ref{20}) can be rewritten as
\begin{equation}
\label{24}
f(r)=1-\frac{2(1+\alpha )r^{\lambda -1}}{(r^{2}+\alpha(1+\alpha ) )^{\frac{\lambda }{2}}}+\frac{\alpha (1+\alpha )r^{\beta  -2}}{(r^{2}+\alpha(1+\alpha ))^{\frac{\beta }{2}}}.
\end{equation}
In  Figs.~1, 2 and 3, we plot the graphs of the metric function $f(r)$ with respect to $r$ for different values of $\alpha$, $\lambda$  and  $\beta$, respectively.  

\begin{figure}[H]
\centering
\begin{minipage}[t]{0.6\linewidth}
\centering
\includegraphics[width=100mm]{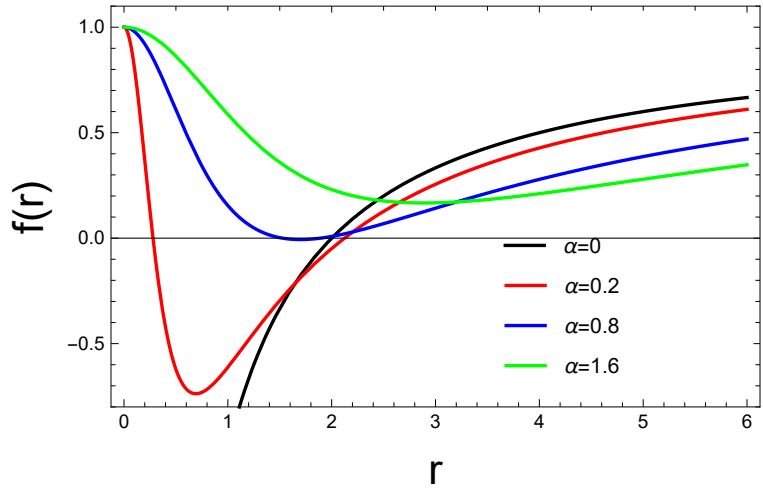}
\caption*{Fig.1. Graph of function $f(r)$ with respect to $r$ for different values of $\alpha$. Here we  choose  $\lambda=3$  and  $\beta=6$.}
\label{fig24} 
\end{minipage}
\end{figure}

\begin{figure}[H]
\centering
\begin{minipage}[t]{0.6\linewidth}
\centering
\includegraphics[width=100mm]{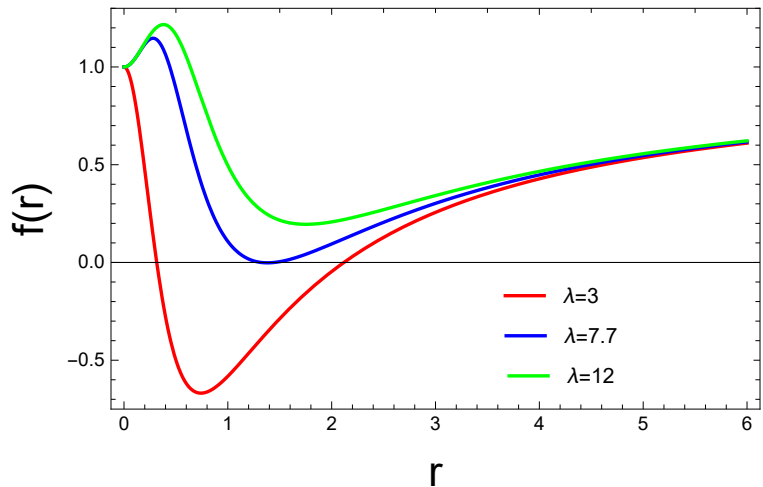}
\caption*{Fig.2. Graph of function $f(r)$ with respect to $r$ for different values of $\lambda$. Here we  choose  $\alpha=0.2$  and  $\beta=4$.}
\label{fig24} 
\end{minipage}
\end{figure}

\begin{figure}[H]
\centering
\begin{minipage}[t]{0.6\linewidth}
\centering
\includegraphics[width=100mm]{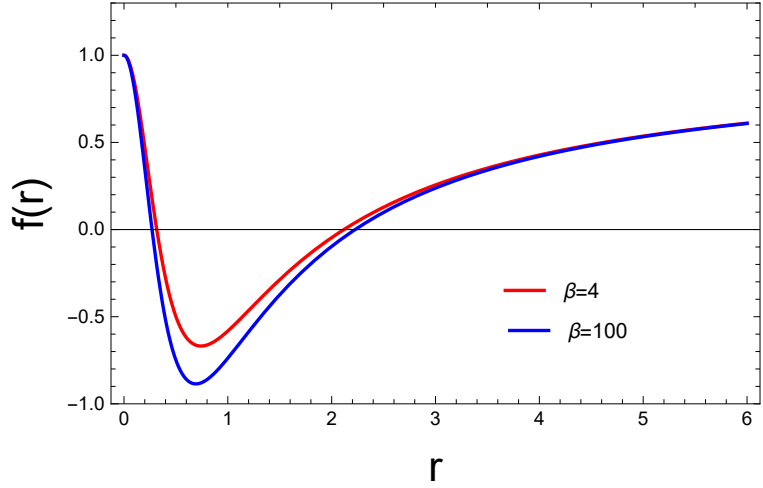}
\caption*{Fig.3. Graph of function $f(r)$ with respect to $r$ for different values of $\beta$. Here we  choose  $\alpha=0.2$  and  $\lambda=3$.}
\label{fig24} 
\end{minipage}
\end{figure}

It should be noted that the black curve  in Fig. 1 represents the case in Einstein's gravity, which is given as a contrast. From Figs.~1 and 2, we can see that the number of the horizons of the generalized ABG STVG black hole will shrink from two to one or even to none when the parameters $\alpha$ and $\lambda$ increase. However, from Fig. 3, we can find that for given $\alpha=0.2$ and  $\lambda=3$, the generalized ABG STVG black hole has two horizons when $\beta$ is set to be 4 and 100, two values with such a big difference. This means that the generalized ABG STVG black hole always maintains two horizons for any $\beta \geqslant 4$.

\section{Quasinormal mode frequencies of gravitational perturbation of the generalized ABG STVG black hole}

The gravitational perturbation of black holes was first performed by Regge and Wheeler~\cite{P4} for the odd parity type of the spherical harmonics, and  then extended to the even parity type by Zerilli~\cite{P5}.  Now we derive the master equation of the gravitational perturbation of the generalized ABG STVG black hole for the odd parity type of the spherical harmonics.

We use $g_{\mu \nu }$ for the background metric and $h_{\mu \nu }$ for the perturbation which is very small compared to $g_{\mu \nu }$. We can calculate  $R_{\mu \nu }$ by  $g_{\mu \nu }$ and  $R_{\mu \nu }+\delta R_{\mu \nu }$ by  $g_{\mu \nu }+\delta g_{\mu \nu }$. Therefore, the expression of $\delta R_{\mu \nu }$ is as follows,
\begin{equation}
\label{25}
\delta R_{\mu \nu }=\nabla _{\rho }\delta \Gamma _{\mu \nu}^{\rho}-\nabla _{\nu  }\delta \Gamma _{\rho  \mu }^{\rho},
\end{equation}
where
\begin{equation}
\label{26}
\delta \Gamma _{ \mu \nu }^{\sigma }=\frac{1}{2}g^{\sigma \gamma }(\nabla_{\mu }h_{\nu \gamma }+\nabla_{\nu  }h_{ \gamma \mu  }-\nabla_{\gamma  }h_{\mu \nu }).
\end{equation}

The canonical form of the  perturbation   in the Regge-Wheeler gauge for the odd parity type of the spherical harmonics takes~\cite{P4} the form,
\begin{equation}
\label{27}
h_{\mu\nu}=\begin{vmatrix}
0 & 0 &  0 &  h_{0}(r) &\\ 
0 & 0 & 0 &  h_{1}(r) &\\ 
0 & 0 &  0 &  0 & \\ 
 h_{0}(r) &  h_{1}(r) & 0 & 0 &
\end{vmatrix}\mathrm{exp}(-i\omega t)\mathrm{sin\theta }\frac{\partial }{\partial \theta }P_{l}(\mathrm{cos\theta}),
\end{equation}
where $\omega$ denotes the complex quasinormal mode frequency, $l$ the  multipole numbers, $P_{l}(\mathrm{cos\theta})$ the Legendre function, and $h_{0}(r)$ and $h_{1}(r)$ two independent components of $h_{\mu \nu }$.

Using Eq.~(\ref{15}), we can obtain the perturbed STVG field equation,
\begin{equation}
\label{28}
\delta G_{\mu \nu }=\delta R_{\mu \nu }-\frac{1}{2}h_{\mu \nu }R-\frac{1}{2}g_{\mu \nu }g^{\rho \gamma }g^{\sigma \tau }h_{\gamma \tau }R_{\rho \sigma }=-2(1+\alpha)h_{\mu \nu }L(P).
\end{equation}
By using the line element Eq.~(\ref{10}) together with Eqs.~(\ref{25}) and (\ref{28}), we can get the following equations,
\begin{equation}
\label{29}
h''_{0}+i\omega h'_{1}+i\omega\frac{2h_{1}}{r}-\frac{h_{0}\left(l^{2}+l-2+2f(r)+(r^{2}f'(r))'+4(1+\alpha )r^{2}L(P)\right)}{r^{2}f(r)}=0,
\end{equation}
\begin{equation}
\label{30}
i\omega h'_{0}-i\omega \frac{2h_{0}}{r}-\omega ^{2}h_{1}+\frac{h_{1}f(r)\left(l^{2}+l-2+(r^{2}f'(r))'+4(1+\alpha )r^{2}L(P)\right)}{r^{2}}=0,
\end{equation}
\begin{equation}
\label{31}
i\omega h_{0}=-f(r)(h_{1}f(r))',
\end{equation}
where the prime represents the derivative with respect to $r$ and $f(r)$ comes from the metric function Eq.~(\ref{24}). By substituting Eq.~(\ref{31}) into Eq.~(\ref{30}), we can eliminate $ h_{0}$. Then defining $\Psi (r)=\frac{f(r)h_{1}(r)}{r}$ and $dr_{*}=\frac{dr}{f(r)}$, we can finally get the master equation,
\begin{equation}
\label{32}
\frac{\mathrm{d}^{2} \Psi(r) }{\mathrm{d} r_{*}^{2}}+[\omega ^{2}-V(r)]\Psi(r) =0,
\end{equation}
where the effective potential $V(r)$ reads
\begin{equation}
\label{33}
V(r)=f(r)\left(\frac{l^{2}+l+2(f(r)-1)+r(rf'(r))'}{r^{2}}+4(1+\alpha )L(P)\right),
\end{equation}
with the nonlinear kinetic term, $L(P)$, for the generalized ABG STVG black hole as follows,
\begin{eqnarray}
\label{34}
L(P)&=&2PH_{P}-H(P)\nonumber \\
&=&\frac{\lambda [-3\alpha (1+\alpha)r^{\lambda -1}+\alpha^{2} (1+\alpha )^{2}(\lambda -1)r^{\lambda -3}]}{2(r^{2}+\alpha +\alpha ^{2})^{\frac{\lambda }{2}+2}}\nonumber \\
& &-\frac{2\alpha r^{\beta }-(5\beta -4)\alpha ^{2}(1+\alpha )r^{\beta -2}+\alpha ^{3}(1+\alpha )^{2}(\beta -1)(\beta -2)r^{\beta -4}}{4(r^{2}+\alpha +\alpha ^{2})^{\frac{\beta }{2}+2}}.
\end{eqnarray}

\subsection{Quasinormal mode frequencies of gravitational perturbation calculated by the 6th order WKB approximation}
Now we use the WKB approximation method to numerically calculate the quasinormal mode frequencies of the gravitational perturbation for the generalized ABG STVG black hole. As to the WKB method, it was first applied to the scattering problem around black holes by Schutz and Will~\cite{P6}. Later, it was  developed to the 3rd order WKB approximation by Iyer and Will~\cite{P7}, to the 6th order by Konoplya~\cite{P8} and most recently to the 13th order by Matyjasek and Opala~\cite{P9}. In order to  study the influence of the parameters $\alpha$, $\lambda$ and $\beta$ on the quasinormal frequencies, we adopt the 6th order WKB approximation for the sake of efficiency in computing the quasinormal frequencies  and focus on the fundamental mode with $l=2$ and $n=0$ due to its dominant ingredient of gravitational waves. The formula of the complex frequency $\omega $ in the 6th order WKB approximation takes~\cite{P8} the form,
\begin{equation}
\label{35}
i\frac{(\omega ^{2}-V_{0})}{\sqrt{-2V^{''}_{0}}}-\sum_{i=2}^{6}\Lambda_{i}=n+\frac{1}{2},
\end{equation}
where $V_{0}$ is the maximum of the effective potential $V(r)$, $V^{''}_{0}=\left.\frac{\mathrm{d^{2}} V(r_{*})}{\mathrm{d} r_{*}^{2}}\right |_{r_{*}=r_{0}}$, $r_{0}$  is the position of the peak value of the effective potential, and 
$\Lambda_{i}$'s are 2nd to 6th order WKB corrections that have been given in Refs.~\cite{P6,P7,P8}.

The numerical results are shown in Tables 1-3.  Correspondingly, we plot the graphs of real parts and negative imaginary parts of quasinormal frequencies with respect to the parameter $\alpha $ in Fig. 4, to the parameter $\lambda $ in Fig. 5, and to the parameter $\beta$ in Fig. 6.

\begin{table}[H]  
\centering 
\caption{The quasinormal mode frequencies of the gravitational perturbation in the generalized ABG STVG black hole, where $\alpha=0$ corresponds to the case of Einstein's gravity.} 
\label{table11} 
\begin{tabular}{|c|c|c|c|c|c|c|c|c|}
\hline
  \multicolumn{2}{|c|}{$M=1$, $l=2$, $n=0$, $\lambda=3$, $\beta=6$}  \\ \hline

$\alpha $ &  $\omega$            \\  \hline
 0  &    0.373619 $-$ 0.088891$i$	 \\ \hline
0.1 &    0.351567 $-$ 0.0802791$i$	 \\ \hline
0.2 &	   0.332646 $-$ 0.0729061$i$	 \\ \hline
0.3 &    0.31627 $-$ 0.0664525$i$ \\ \hline	
0.4 &    0.302005 $-$ 0.060682$i$ \\ \hline	
0.5 &	    0.289524 $-$ 0.0554096$i$	 \\ \hline	
0.6 &	    0.278565 $-$ 0.0504797$i$	 \\ \hline
0.7 &	    0.268911 $-$ 0.0457465$i$	 \\ \hline
0.8 &	    0.260347 $-$ 0.0410612$i$	 \\ \hline	
\end{tabular}
\end{table}

\begin{table}[H]  
\centering 
\caption{The quasinormal mode frequencies of the gravitational perturbation in the generalized ABG STVG black hole, where $\lambda =3$ corresponds to the case of the ABG STVG black hole.} 
\label{table13} 
\begin{tabular}{|c|c|c|c|c|c|c|c|c|}
\hline
  \multicolumn{2}{|c|}{$M=1$, $l=2$, $n=0$, $\alpha=0.2$, $\beta=4$}  \\ \hline

$\lambda $ &  $\omega$            \\  \hline
3	& 0.332935 $-$ 0.0727599$i$	  \\  \hline
3.5	& 0.335287 $-$ 0.0722335$i$	  \\  \hline
4	& 0.33774 $-$ 0.0716394$i$	  \\  \hline
4.5	& 0.340303 $-$ 0.0709636$i$	  \\  \hline
5	& 0.342984 $-$ 0.0701877$i$	  \\  \hline
5.5	& 0.345796 $-$ 0.069288$i$	  \\  \hline
6	& 0.348749 $-$ 0.0682323$i$	   \\  \hline	
6.5	& 0.351846 $-$ 0.0669783$i$	  \\  \hline
7	& 0.355089 $-$ 0.0654691$i$	  \\  \hline
7.5	& 0.358457 $-$ 0.0636334$i$	  \\  \hline
7.7	& 0.359829 $-$ 0.0627905$i$	  \\  \hline

\end{tabular}
\end{table}

\begin{table}[H]  
\centering 
\caption{The quasinormal mode frequencies of the gravitational perturbation in the generalized ABG STVG black hole, where $\beta=4$ corresponds to the case of the ABG STVG black hole.} 
\label{table14} 
\begin{tabular}{|c|c|c|c|c|c|c|c|c|}
\hline
  \multicolumn{2}{|c|}{$M=1$, $l=2$, $n=0$, $\alpha=0.2$, $\lambda=3$}  \\ \hline

$\beta $ &  $\omega$            \\  \hline
4 	&0.332935  $-$ 0.0727599$i$  \\  \hline	
5	&0.332789  $-$ 0.0728348$i$	\\  \hline	
6	&0.332646  $-$ 0.0729061$i$	\\  \hline	
7	&0.332505  $-$ 0.0729737$i$	\\  \hline	
8	&0.332367  $-$ 0.073038$i$	\\  \hline	
9	&0.332231  $-$ 0.0730992$i$	\\  \hline	
10	&0.332096  $-$ 0.0731574$i$	\\  \hline	
11	&0.331965  $-$ 0.0732124$i$	\\  \hline	
12	&0.331835  $-$ 0.0732647$i$	\\  \hline	

\end{tabular}
\end{table}

\begin{figure}[h]
		\centering
		\begin{minipage}{.5\textwidth}
			\centering
			\includegraphics[width=80mm]{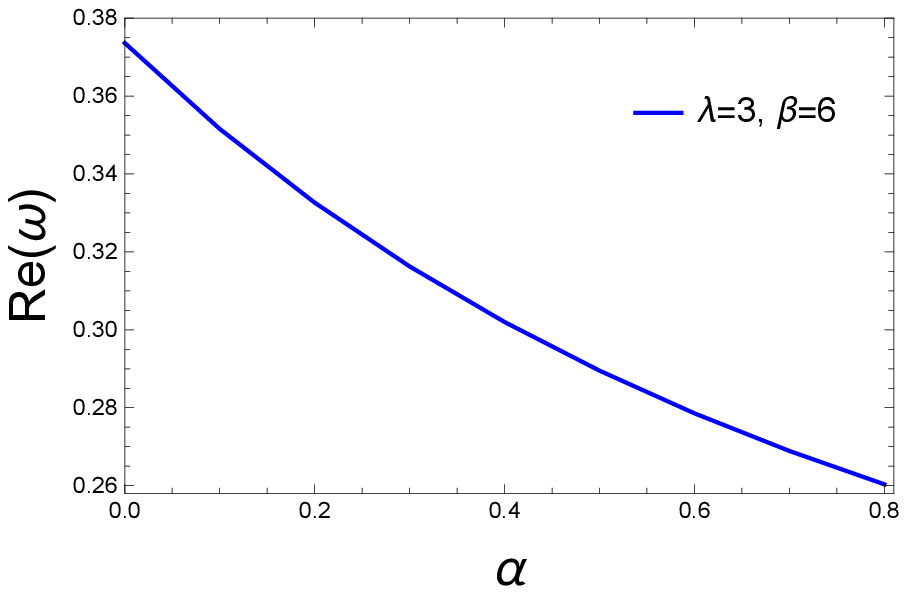}
		\end{minipage}%
		\begin{minipage}{.5\textwidth}
			\centering
			\includegraphics[width=80mm]{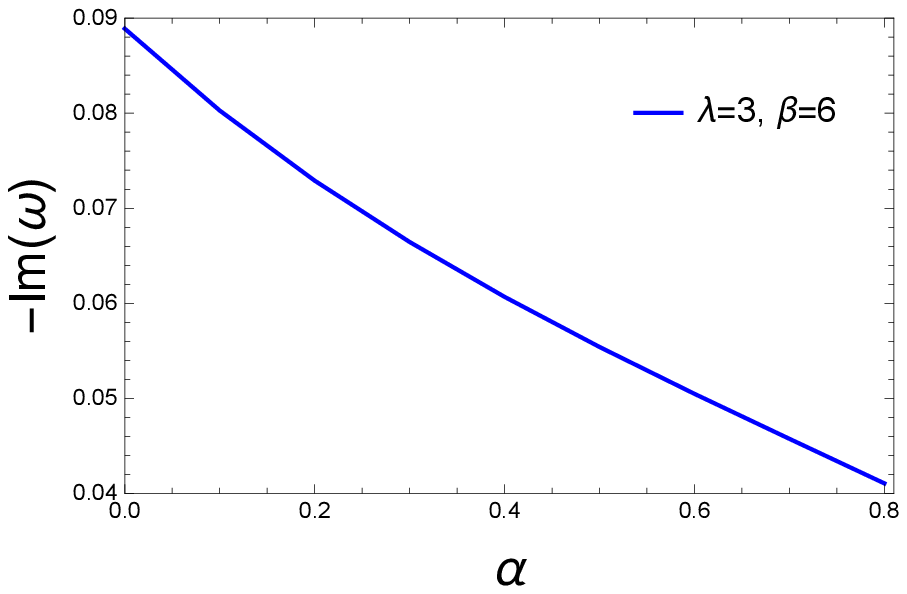}
		\end{minipage}
		\caption*{Fig. 4. Graph of real parts and negative imaginary parts  of quasinormal  frequencies of    the gravitational perturbation with respect to  $\alpha $. Here we choose $M=1$, $l = 2$ and $n=0$.}\label{figure4}
\end{figure}

\begin{figure}[h]
		\centering
		\begin{minipage}{.5\textwidth}
			\centering
			\includegraphics[width=80mm]{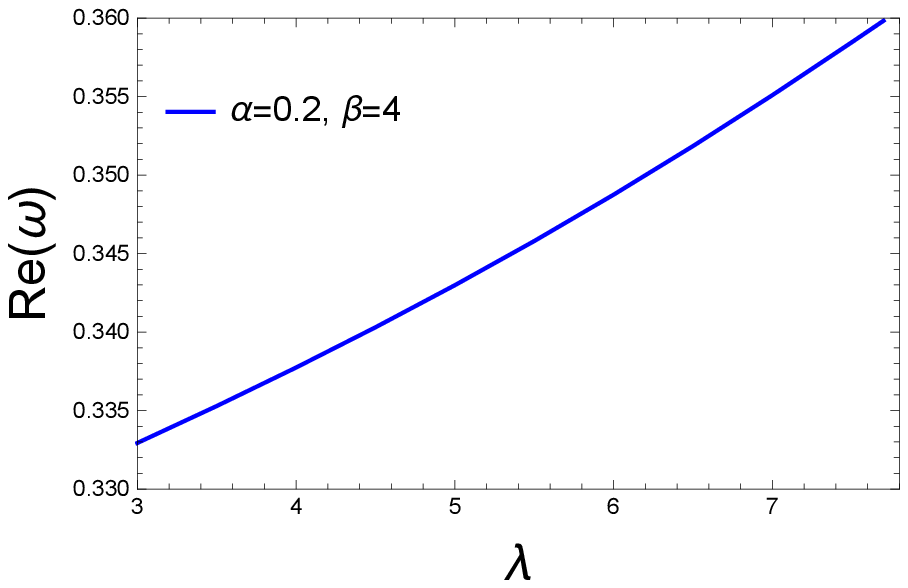}
		\end{minipage}%
		\begin{minipage}{.5\textwidth}
			\centering
			\includegraphics[width=80mm]{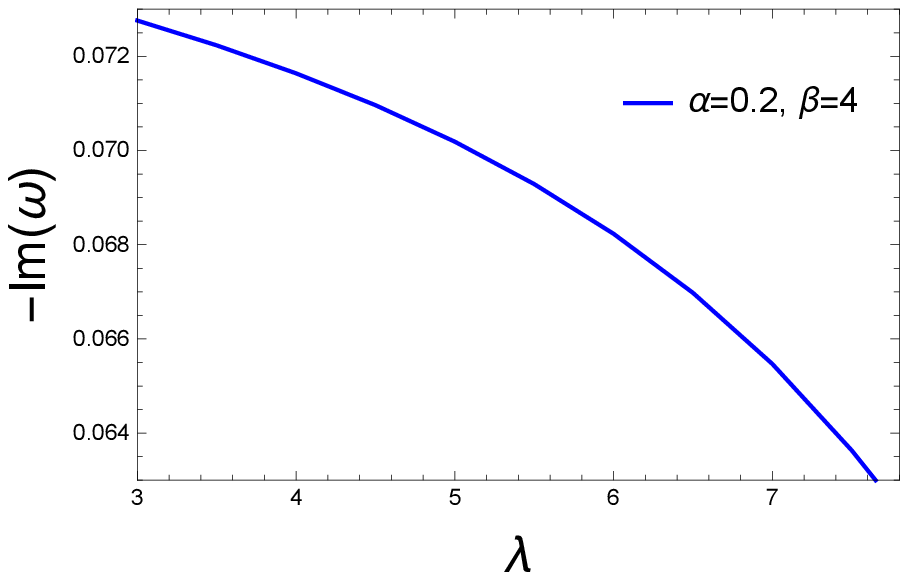}
		\end{minipage}
		\caption*{Fig. 5. Graph of real parts and negative imaginary parts  of quasinormal  frequencies of    the gravitational perturbation with respect to  $\lambda $. Here we choose $M=1$, $l = 2$ and $n=0$.}\label{figure5}
\end{figure}

\begin{figure}[h]
		\centering
		\begin{minipage}{.5\textwidth}
			\centering
			\includegraphics[width=80mm]{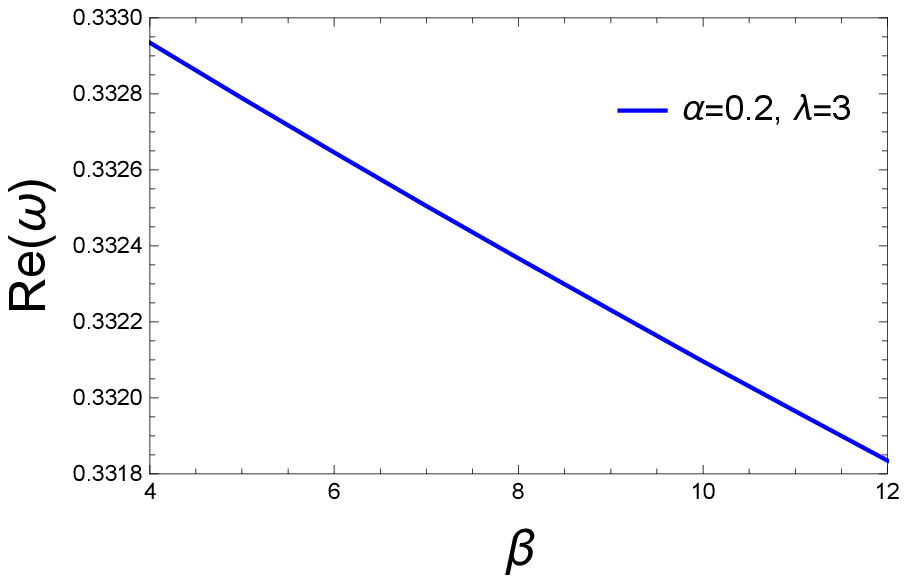}
		\end{minipage}%
		\begin{minipage}{.5\textwidth}
			\centering
			\includegraphics[width=80mm]{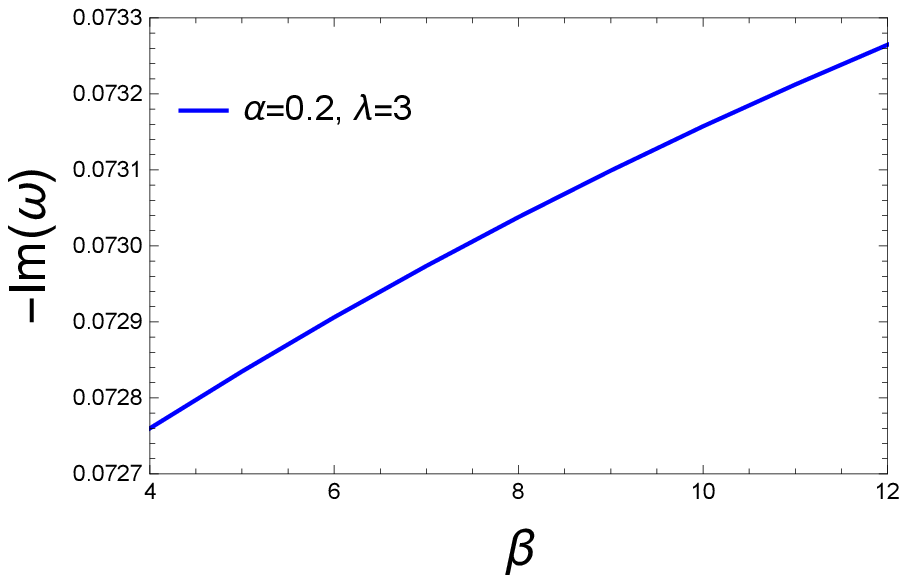}
		\end{minipage}
		\caption*{Fig. 6. Graph of real parts and negative imaginary parts  of quasinormal  frequencies of    the gravitational perturbation with respect to  $\beta$. Here we choose $M=1$, $l = 2$ and $n=0$.}\label{figure6}
\end{figure}

From Fig. 4 for the  given values of $\lambda$ and $\beta$, we can see that the increase of the parameter $\alpha$ makes the real part and the absolute value of the imaginary part of quasinormal frequencies decrease. From Fig. 5 for the given values of $\alpha$ and $\beta$, we can see that the increase of the parameter $\lambda$ makes the real part of quasinormal frequencies increase but the absolute value of the imaginary part decrease. From Fig. 6  for the  given values of $\alpha$ and $\lambda$, we can see that the increase of the parameter $\beta$ makes the real part of quasinormal frequencies decrease but the absolute value of the imaginary part increase.

\subsection{Quasinormal mode frequencies at the eikonal limit calculated via circular null geodesics}

The circular null geodesic method was first proposed by Cardoso et al.~\cite{P10} for the calculation of quasinormal mode frequencies of a static spherically symmetric black hole at the eikonal limit, $l\gg 1$. Under this limit, the effective potential $V(r)$, Eq.~(\ref{33}), goes to the following form,
\begin{equation}
\label{36}
V(r)=f(r)\frac{l^{2}}{r^{2}}+O(l).
\end{equation}
The quasinormal mode frequencies $\omega $ can be expressed as follows,
\begin{equation}
\label{37}
\omega _{l\gg 1}=l\Omega_c -i\left(n+\frac{1}{2}\right)|\lambda_{\rm L}|,
\end{equation}
where the angular velocity $\Omega_c $ and Lyapunov exponent $\lambda_{\rm L}$ determine the real and imaginary parts of the quasinormal mode frequency $\omega $, respectively,\begin{equation}
\label{38}
\Omega _c=\frac{\sqrt{f(r_{c})}}{r_{c}},        \qquad   \lambda_{\rm L} =\sqrt{\frac{f(r_{c})(2f(r_{c})-r_{c}^{2}{f}''(r_{c}))}{2r_{c}^{2}}},
\end{equation}
with $r_{c}$ the radius of circular null geodesics determined by
\begin{equation}
\label{39}
2f(r_{c})-r_{c}\left.\frac{\mathrm{d} f(r)}{\mathrm{d} r}\right|_{r=r_{c}}=0.
\end{equation}
We note that Eq.~(\ref{37}) is not a universal feature for static spherically symmetric black holes in an asymptotically flat spacetime with arbitrary dimensions. For example, the electromagnetic perturbation in the general relativity coupled to nonlinear electrodynamics does not satisfy~\cite{P37,P38} this relation, and the gravitational perturbation in the Einstein-Lovelock gravity does not~\cite{P39,P40}, either. Nonetheless, we emphasize that Eq.~(\ref{37}) keeps valid in the gravitational perturbation for the generalized ABG STVG black hole because its effective potential Eq.~(\ref{33}) matches the form given by Ref.~\cite{P41}. 

Recently, the relationship between the real part of quasinormal frequencies at the eikonal limit and the black hole shadow radius has been obtained~\cite{P35}, i.e., ${\rm Re}\,\omega _{l\gg 1}=\frac{l}{R_{S}}$, where  $R_{S}=\frac{r_{c}}{\sqrt{f(r_{c})}}$ is the shadow radius of black holes and is equal to the inverse of the angular velocity $\Omega_c $ at the eikonal limit. Subsequently, Cuadros-Melgar et al. proposed~\cite{P36} an improved relation between the real part of quasinormal frequencies at the eikonal limit and the shadow radius by using the WKB method, i.e., ${\rm Re}\,\omega _{l\gg 1}=\frac{l+\frac{1}{2}}{R_{S}}$.  Therefore, the improved expression of the quasinormal mode frequencies at the eikonal limit takes the form,
\begin{equation}
\label{40}
 \omega _{l\gg 1}=\frac{l+\frac{1}{2}}{R_{S}}-i\left(n+\frac{1}{2}\right)|\lambda_{\rm L}|=\left(l+\frac{1}{2}\right)\Omega_c -i\left(n+\frac{1}{2}\right)|\lambda_{\rm L}|.
\end{equation}

In order to test the above relation in the generalized ABG STVG black hole spacetime, we use the 6th order WKB approximation method on the one hand, and on the other hand the improved expression Eq.~(\ref{40}), and then we compare the results computed by the two ways.
We focus on the fundamental modes of the odd parity gravitational perturbation for different multipole numbers $l$, as shown in Tables 4 and 5, where $\omega _{\rm R}$ and $-\omega _{\rm I}$ are the real parts and negative imaginary parts of quasinormal  frequencies calculated by the 6th order WKB approximation method,  and $\omega _{\rm {R1}}=l\Omega_c $  and $\omega _{\rm {R2}}=\frac{l+\frac{1}{2}}{R_{S}}=(l+\frac{1}{2})\Omega_c $ are the real parts calculated by the formulas Eqs.~(\ref{37}) and (\ref{40}),  respectively, and $\omega _{\rm Ii}=\frac{1}{2}|\lambda_{\rm L}|$, ${\rm i}=1, 2$, are determined by the Lyapunov exponent $\lambda_{\rm L}$ at the eikonal limit.

\begin{table}[htbp]
\centering
\caption{The real part of quasinormal frequencies of the gravitational perturbation in the generalized ABG STVG black hole  for different multipole numbers $l$, where $M=1$, $\lambda=3$, $\beta=6$ and $n=0$.}
\begin{tabular}{|c|c|c|c|c|c|c|}
\hline
  \multicolumn{1}{|c|}{$l$}   &   \multicolumn{1}{c|}{$\alpha$}  &    \multicolumn{1}{c|}{$\omega _{\rm R}$ }   & \multicolumn{1}{c|}{$\omega _{\rm R1}$}  & \multicolumn{1}{c|}{$\omega _{\rm R2}$}  &   \multicolumn{1}{c|}{$\left | \frac{\omega _{\rm R1}-\omega _{\rm R}}{\omega _{\rm R}} \right |$}   &   \multicolumn{1}{c|}{$\left | \frac{\omega _{\rm R2}-\omega _{\rm R}}{\omega _{\rm R}} \right |$} \\
\hline
  \multirow{4}[0]{*}{10}   
&0 &      1.996788 & 1.924501&	2.020726&	0.03620164& 	0.011988253     \\  \cline{2-7}
&0.2& 	1.770147&	1.705669&	1.790952&	0.036425223&	0.011753261  \\    \cline{2-7} 
&0.5&	      1.534037&	1.477736&	1.551623&	0.036701201&	0.01146387    \\    \cline{2-7}
&0.8& 	1.376202&	1.325461&	1.391735&	0.036870314&	0.011286861   \\    \cline{2-7}
\hline 
\multirow{4}[0]{*}{100}   

&0&	 19.338743&	     19.245009&  19.341234&     0.004846954& 	0.000128809  \\  \cline{2-7}
&0.2&  17.139805&	17.056687&	 17.14197& 	0.004849413&	0.000126314  \\  \cline{2-7}
&0.5&	 14.849422& 	14.777365&	 14.851252&	0.004852512&	0.000123237   \\  \cline{2-7}
&0.8&	 13.319273&  	13.254615&	 13.320888&	0.004854469&	0.000121253   \\  \cline{2-7}

\hline 
\multirow{4}[0]{*}{1000}   

&0&	192.546065&	192.45009& 	192.546315&	0.000498452&	$1.29839\times 10^{-6}$   \\  \cline{2-7}
&0.2&	170.651933&	170.566867&	170.65215& 	0.000498477&	$1.27159\times 10^{-6}$   \\  \cline{2-7}
&0.5&	147.847351&	147.773648&	147.847535&	0.000498507&	$1.24453\times 10^{-6}$   \\  \cline{2-7}
&0.8&	132.612261&	132.54615& 	132.612423&	0.000498529&	$1.22161\times 10^{-6}$   \\  \cline{2-7}

\hline 
\multirow{4}[0]{*}{100000}   

&0&	19245.105196&	  19245.008973&	19245.105198&  	$4.99987\times 10^{-6}$ &        	$1.03923\times 10^{-10}$  \\  \cline{2-7}
&0.2&	 17056.771963&	 17056.686681&	17056.771965&	$4.99989\times 10^{-6}$ &       	$1.17256\times 10^{-10}$  \\  \cline{2-7}
&0.5&	 14777.438705&	 14777.36482&	14777.438707&	$4.99985\times 10^{-6} $&       	$1.35341\times 10^{-10}$  \\  \cline{2-7}
&0.8&	 13254.681241&	13254.61497&	13254.681243&	$4.99982\times 10^{-6}$ &       	$1.5089\times 10^{-10}$   \\  \cline{2-7}

\hline
  \end{tabular}%
  \label{tab:addlabel}%
\end{table}%

\begin{table}[htbp]
\centering
\caption{The negative  imaginary parts  of quasinormal frequencies of the gravitational perturbation in the generalized ABG STVG black hole  for different multipole numbers $l$, where $M=1$, $\lambda=3$, $\beta=6$ and $n=0$.}
\begin{tabular}{|c|c|c|c|c|}
\hline
  \multicolumn{1}{|c|}{$l$}   &   \multicolumn{1}{c|}{$\alpha$}  &   \multicolumn{1}{c|}{$-\omega _{\rm I}$}   & \multicolumn{1}{c|}{$-\omega _{\rm Ii}$}    &   \multicolumn{1}{c|}{$\left | \frac{\omega _{\rm Ii}-\omega _{\rm I}}{\omega _{\rm I}} \right |$}    \\
\hline
  \multirow{4}[0]{*}{10}   

&0&	     0.0958639& 	0.096225&  	0.003766799  \\  \cline{2-5}
&0.2& 	0.0785097& 	0.0787893& 	0.003561343  \\  \cline{2-5}
&0.5& 	0.0595306& 	0.0597276& 	0.003309222  \\  \cline{2-5}
&0.8& 	0.0442768& 	0.0444174& 	0.003175478  \\  \cline{2-5}

\hline
  \multirow{4}[0]{*}{100}   

&0&	      0.0962211&	0.096225&	     $4.05317\times 10^{-5}$       \\  \cline{2-5}
&0.2& 	0.0787862& 	0.0787893& 	$3.9347\times 10^{-5}$       \\  \cline{2-5}
&0.5& 	0.0597255& 	0.0597276& 	$3.51609\times 10^{-5}$      \\  \cline{2-5}
&0.8& 	0.0444159& 	0.0444174& 	$3.37717\times 10^{-5}$       \\  \cline{2-5}

\hline
  \multirow{4}[0]{*}{1000}

&0&	      0.096225& 	0.096225&	      0                                  \\  \cline{2-5}
&0.2& 	0.0787892& 	0.0787893& 	$1.26921\times 10^{-6}$            \\  \cline{2-5}
&0.5& 	0.0597276& 	0.0597276& 	0                                  \\  \cline{2-5}
&0.8& 	0.0444174& 	0.0444174& 	0                                  \\  \cline{2-5}

\hline
  \multirow{4}[0]{*}{100000}  
&0&	     0.096225&	     0.096225&	      0      \\  \cline{2-5}
&0.2& 	0.0787893& 	0.0787893& 	 0      \\  \cline{2-5}
&0.5& 	0.0597276& 	0.0597276&  	0       \\  \cline{2-5}
&0.8& 	0.0444174& 	0.0444174& 	0       \\  \cline{2-5}

\hline
  \end{tabular}%
  \label{tab:addlabel2}%
\end{table}%

From Table 4, we can see that the improved relationship between the real part of quasinormal frequencies at the eikonal limit and the shadow radius is valid for the generalized ABG STVG black hole, and further that $\omega _{\rm {R2}}$ is closer to  $\omega _{\rm {R}}$ than $\omega _{\rm {R1}}$ for a bigger multipole number $l$. That is, we find that $\omega _{\rm {R2}}$ is more accurate than $\omega _{\rm {R1}}$, or in other words, the relative error of $\omega _{\rm {R2}}$ and $\omega _{\rm {R}}$ is less than that of $\omega _{\rm {R1}}$ and $\omega _{\rm {R}}$ for a bigger $l$. In addition, from Table 5, we can see that the imaginary part  of quasinormal frequencies, similar to the real part, is more accurate for a bigger $l$.

By substituting the metric function of the generalized ABG STVG black hole, Eq.~(\ref{24}), into Eqs.~(\ref{38}) and ~(\ref{39}), we obtain the angular velocity $\Omega_c $ and Lyapunov exponent $\lambda_{\rm L}$. Then we can plot the graphs of $\Omega_c $ versus $\lambda_{\rm L}$ with respect to the parameters $\alpha$, $\lambda$ and $\beta$ in Figs. 7--9, respectively.

\begin{figure}[h]
		\centering
		\begin{minipage}{.5\textwidth}
			\centering
			\includegraphics[width=80mm]{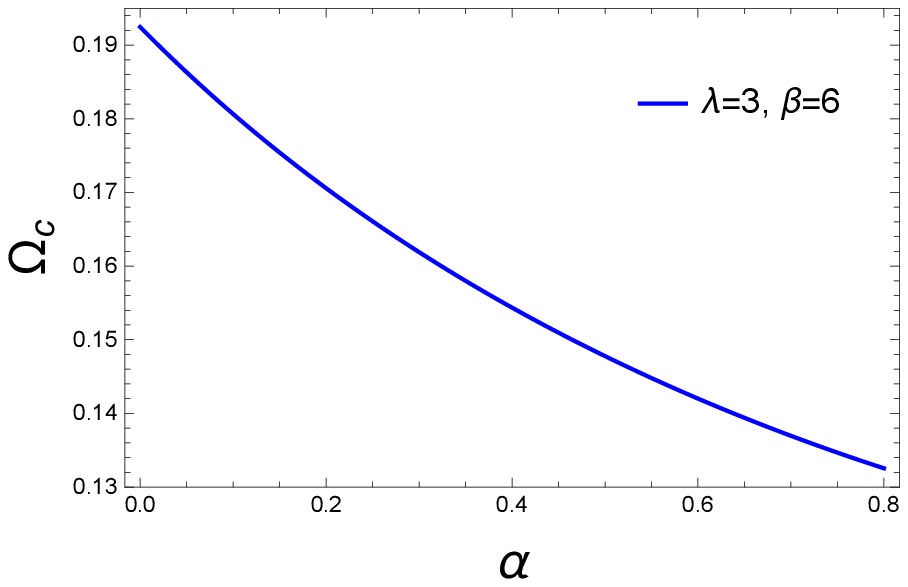}
		\end{minipage}%
		\begin{minipage}{.5\textwidth}
			\centering
			\includegraphics[width=80mm]{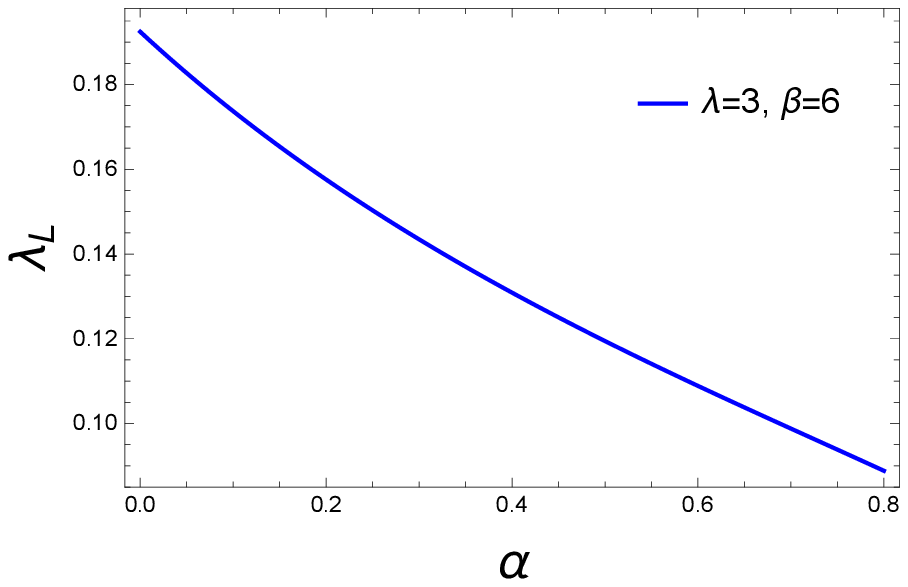}
		\end{minipage}
		\caption*{Fig. 7. Graph of the angular velocity $\Omega_c$  and the Lyapunov exponent $\lambda_{\rm L}$  with respect  to  the parameter $\alpha $. Here we choose $M=1$.}
\label{figure7}
\end{figure}

\begin{figure}[h]
		\centering
		\begin{minipage}{.5\textwidth}
			\centering
			\includegraphics[width=80mm]{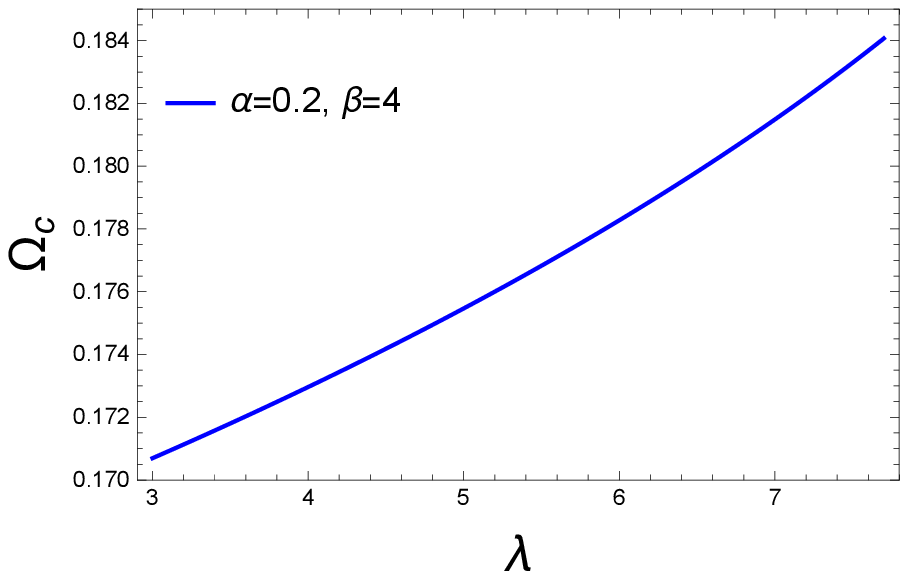}
		\end{minipage}%
		\begin{minipage}{.5\textwidth}
			\centering
			\includegraphics[width=80mm]{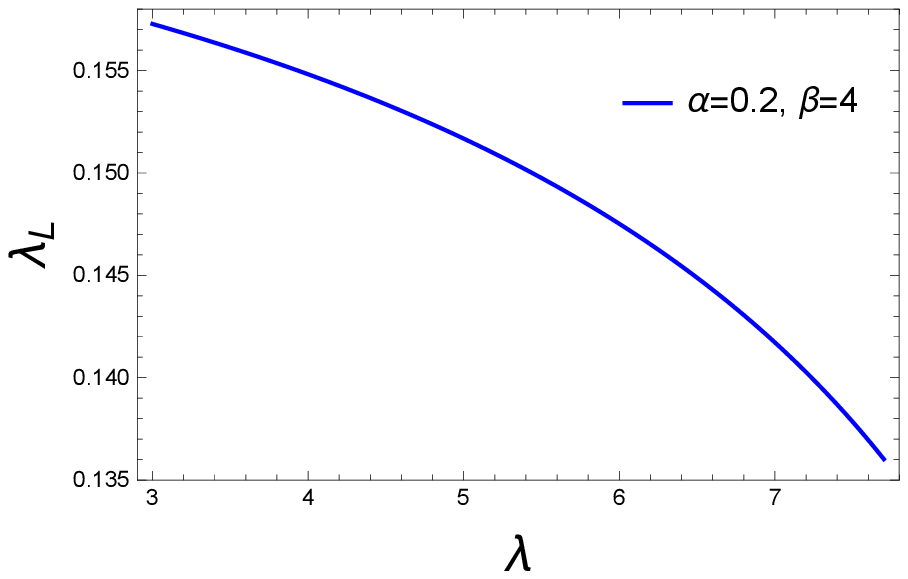}
		\end{minipage}
		\caption*{Fig. 8. Graph of the angular velocity $\Omega_c$  and the Lyapunov exponent $\lambda_{\rm L}$  with respect  to  the parameter $\lambda$. Here we choose $M=1$.}
\label{figure8}
\end{figure}

\begin{figure}[h]
		\centering
		\begin{minipage}{.5\textwidth}
			\centering
			\includegraphics[width=80mm]{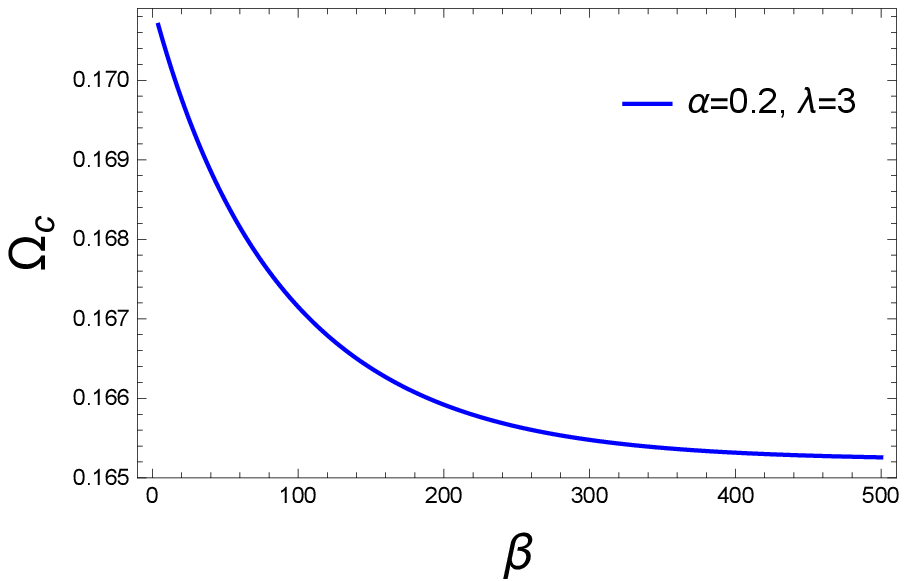}
		\end{minipage}%
		\begin{minipage}{.5\textwidth}
			\centering
			\includegraphics[width=80mm]{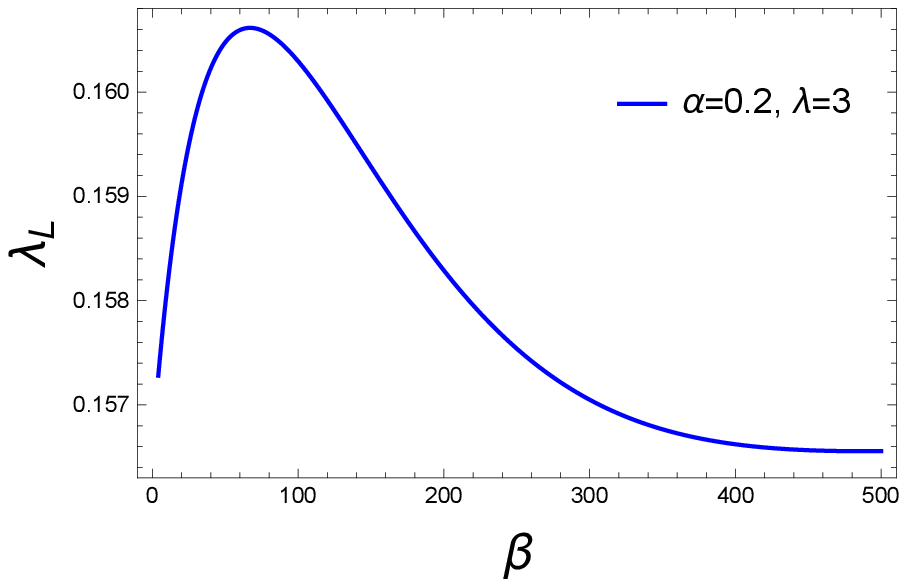}
		\end{minipage}
		\caption*{Fig. 9. Graph of the angular velocity $\Omega_c$  and the Lyapunov exponent $\lambda_{\rm L}$  with respect  to  the parameter $\beta$. Here we choose $M=1$.}
\label{figure9}
\end{figure}

From Fig. 7 for the  given values of $\lambda$ and $\beta$, we can see that the increase of the parameter $\alpha$ makes the angular velocity $\Omega_c$ and the Lyapunov exponent $\lambda_{\rm L}$ decrease. From Fig. 8 for the  given values of $\alpha$ and $\beta$, we can see that the increase of the parameter $\lambda$ makes the angular velocity $\Omega_c$ increase but the Lyapunov exponent $\lambda_{\rm L}$ decrease. From Fig. 9 for the  given values of $\alpha$ and $\lambda$, we can see that  the increase of the parameter $\beta$ makes  the angular velocity $\Omega_c$  decrease monotonically but the Lyapunov exponent  $\lambda_{\rm L}$ increase to a maximum at first and then decrease. It should be noted that the difference between Fig. 9 and Fig. 6 is caused by the great difference in the range of values of the parameter $\beta$.

Through the comparison and analysis of Figs. 4--6  and   Figs. 7--9, we can finally get the following conclusions:
\begin{itemize}
\item  The larger the parameter $\alpha$ is, the more slowly the gravitational wave oscillates and decays in the generalized ABG STVG black hole spacetime.
\item  The larger the parameter  $\lambda$ is, the faster the gravitational wave oscillates but the more slowly it decays in the generalized ABG STVG black hole spacetime.
\item  The larger the parameter $\beta$ is, the more slowly the gravitational wave oscillates, but it decays faster at first and then more slowly in the generalized ABG STVG black hole spacetime. In particular, there exists a special value of $\beta$ that makes the gravitational wave decay fastest in the generalized ABG STVG black hole spacetime.

\end{itemize}

Finally, we make a test on the stability of the generalized ABG STVG black hole under linear perturbations and find that instabilities arise  in the sector of the even parity perturbation. Considering the similarity between the generalized ABG STVG black hole and the generalized ABG  black hole~\cite{P3} in Einstein's gravity, we can directly use the stability conditions which are suitable~\cite{P43} to the two kinds of regular black holes under linear perturbations. The stability conditions are as follows:
\begin{equation}
\label{42}
H(P)<0, \qquad  H_{P}>0, \qquad  0<f(r)Y\leq 3, 
\end{equation}
for any $r>r_{\rm H}$.
Here $r_{\rm H}$ is the event horizon of the black hole,  $Y\equiv (1+\frac{2PH_{PP}}{H_{P}})$, and $H_{PP}$ is the second derivative of $H(P)$ with respect to $P$.  In addition, it is pointed out in Ref.~\cite{P43} that if $Y$ is negative in the region outside the event horizon, the black hole is unstable to the linear even parity perturbation with sufficiently large multipole numbers $l$. Considering the complexity of the metric function Eq.~(\ref{24}) of the generalized ABG STVG black hole, we select six sets of data for $(\lambda, \beta)$: $(3, 4)$, $(3, 6)$, $(6, 6)$, $(7, 4)$, $(3, 400)$, and $(6, 300)$, and substitute them into Eq.~(\ref{42}). We seek out the region outside the event horizon of the generalized ABG STVG black hole in the case of $(3, 400)$, such that $Y<0$ is satisfied. This implies that the generalized ABG STVG black hole is unstable to the linear even parity perturbation with sufficiently large multipole numbers $l$ under a certain set of data of $(\lambda, \beta)$.

\section{Conclusion }

In this paper, we first construct the solution of the generalized ABG STVG black hole and analyze the features of the black hole. We find that the generalized ABG STVG black hole is regular when $\lambda \geqslant 3$ and  $\beta \geqslant 4$. We also examine the influence of the parameters $\alpha$, $\lambda$ and $\beta$ on the horizon of the generalized ABG STVG black hole. We derive the master equation of gravitational perturbation of the generalized ABG STVG black hole for the odd parity type of the spherical harmonics and  calculate the quasinormal mode frequencies by using the 6th order WKB approximation method. We draw the graphs of real parts and imaginary parts of quasinormal frequencies of the gravitational perturbation with respect to the parameters $\alpha$, $\lambda$ and $\beta$, respectively. We also use the null geodesic method to compute the quasinormal mode frequencies at the eikonal limit. We finally get the following conclusions: The increase of the parameters $\alpha$ and $\lambda$ in the generalized ABG STVG black hole  spacetime makes the gravitational waves decay slowly, while the increase of the parameter $\beta$ makes the gravitational wave decay fast at first and then slowly. Furthermore, we verify that  the improved correspondence  between  the real part of quasinormal frequencies at the eikonal limit and the black hole shadow radius, ${\rm Re}\,\omega _{l\gg 1}=\frac{l+\frac{1}{2}}{R_{S}}$, is valid for the generalized ABG STVG black hole.
  
At last, we note that the quasinormal modes of the odd parity gravitational perturbation for the black holes with nonlinear electrodynamics are not necessarily equal~\cite{P45} to those of the even parity gravitational perturbation. It has been proved~\cite{P44} that the isospectrality is broken in deformed Reissner-Nordstr\"om black holes, which is also called a parity splitting phenomenon. We thus speculate that the generalized ABG STVG  black hole is very likely to have different  quasinormal frequencies under odd parity and even parity perturbations. Therefore, it is meaningful to calculate the quasinormal mode frequencies of the even parity gravitational perturbation for the generalized ABG STVG black hole in our future work.

\section*{Acknowledgments}

The authors would like to thank  C. Lan, J. Moffat, and H. Yang for helpful discussions. They also thank the anonymous referee very much for the helpful comments that improve this work greatly.
This work was supported in part by the National Natural Science Foundation of China under Grant No. 11675081.


\end{document}